\documentclass[
aps,%
12pt,%
final,%
notitlepage,%
oneside,%
onecolumn,%
nobibnotes,%
nofootinbib,% In the current version of REVTeX, when this option is on,
superscriptaddress,%
noshowpacs,%
centertags]%
{revtex4-1}
\RequirePackage{graphicx}

\def\refitem#1{\relax}

\begin{document}
\title{Confining but chirally symmetric dense and cold matter}

\author{\firstname{L. Ya.} \surname{Glozman}}
\email{leonid.glozman@uni-graz.at}
\affiliation{Institute for  Physics, Theoretical Physics Branch,
 University of Graz, 
A-8010, Graz, Austria}

\begin{abstract}
The possibility for existence of cold, dense
chirally symmetric matter with confinement is reviewed. 
The answer to this question
crucially depends on the mechanism of mass generation in QCD
and interconnection of confinement and chiral symmetry
breaking. This question can be clarified from spectroscopy
of hadrons and their axial properties. Almost systematical 
parity doubling of highly excited hadrons suggests that
their mass is not related to chiral symmetry breaking
in the vacuum and is approximately chirally symmetric. Then
there is a possibility for existence of confining but chirally
symmetric matter. We clarify a possible mechanism underlying
such a phase at low temperatures and large density. 
Namely, at large density the Pauli blocking prevents
the gap equation to generate a solution with broken chiral
symmetry. However, the chirally symmetric part of the quark
Green function as well as all color non-singlet quantities
 are still infrared divergent, meaning that the system is
 with confinement. A possible phase transition to such a matter
 is most probably of the first order. This is because there are
 no  chiral partners to  the lowest lying hadrons.

\end{abstract}

\maketitle

\section{Introduction}

A key question to QCD at high temperatures and densities is whether
and  how deconfinement and chiral restoration transitions (crossovers)
are connected to each other. In order to answer this question we
need understanding of hadron mass generation  in QCD, how both confinement
and chiral symmetry breaking influence the origin of mass. We
know from the trace anomaly in QCD that the hadron mass (we discuss here
only the light quark sector) almost entirely consists of the energy
of quantized gluonic field. However, this tells us nothing about the effect
of chiral symmetry breaking on hadron mass. The chiral symmetry is
dynamically broken in the QCD vacuum and this phenomenon is crucially
important for the mass origin of the lowest lying hadrons, such as
pion, nucleon or rho-meson. Phenomenologically it follows from the
well established (pseudo) Nambu-Goldstone nature of pion as well as
from the absence of chiral partners to the lowest lying hadrons.
 Their mass is determined to large
degree by the quark condensate of the vacuum, which can be seen 
from the SVZ sum rules 
\cite{Shifman:1978bx,Ioffe:1981kw}, as well as from many 
different microscopical models.  

At the same time almost systematical  parity doubling
in both highly excited baryons \cite{G1} and mesons \cite{G2} 
suggests that  chiral symmetry breaking in the vacuum is almost irrelevant
to the mass generation of these hadrons, i.e., the chiral (and $U(1)_A$)
symmetry gets effectively restored, for a review see \cite{G3}. This
conjecture is strongly supported by the pattern of strong decays of
excited hadrons \cite{G4}. Experimental discovery of the
still missing chiral partners to some high lying states is required, however,
for the unambiguous conclusion \cite{GS,KLEMPT}.

If effective chiral restoration in excited hadrons is correct, then
 there is a possibility in QCD for a phase
with confinement (i.e., elementary excitations are of the color-singlet
hadronic type) and where at the same time chiral symmetry is restored.

\section{Quarkyonic matter}

In the large $N_c$ world with quarks in the fundamental representation there
are neither dynamical quark - antiquark nor quark - quark hole loops. Consequently,
at low temperatures there is no Debye screening of the confining 
gluon propagator and  gluodynamics  is
the same as in vacuum. Confinement persists up to arbitrary large 
density. In such  case it is possible to define  quarkyonic matter \cite{McLerran:2007qj}. 
In short,
it is a strongly interacting matter with confinement and with a well defined Fermi
sea of baryons or quarks. At smaller densities it should be a Fermi sea of nucleons
(so it matches with standard nuclear matter), while at higher densities, when  nucleons 
are in a strong overlap, a quark Fermi surface should be formed. While a quark Fermi sea
is formed, the system is still with confinement and excitation modes are of the color-singlet
hadronic type. 
Obviously, the question of existence or nonexistence of the
quarkyonic phase in the real world $N_c =3$ cannot be formulated and 
studied with models
that lack explicit confinement of quarks. 

Note that nothing can be concluded from this simple argument
about existence or nonexistence of the chiral restoration phase
transition, i.e., whether there is or not a "subphase" 
with restored chiral symmetry within the 
quarkyonic matter. (Quite often, unfortunately,
a  quarkyonic
matter is confused with  confining but chirally symmetric phase.)

In the large $N_c$ 't Hooft limit such a matter persists at low temperatures 
up to arbitrary large densities. At
which densities in the real  $N_c=3$ world will we have a deconfining transition (which could be a very
smooth crossover) to a quark matter with uncorrelated single quark  excitations? 
Lattice results for  $N_c=2$ suggest that such
a transition could occur at densities of the order 100*nuclear matter density \cite{Hands:2010gd}.
If correct, then it would imply that at all densities relevant to future experiments
and astrophysics we will have a dense quarkyonic (baryonic) 
matter with confinement.

The most interesting question concerns the fate  of chiral symmetry 
breaking
in this dense, cold quarkyonic matter with confinement. Indeed, in a dense
matter one expects that all lowest lying quark levels (that are required
for the very existence of the quark condensate) are occupied by valence quarks
and  Pauli blocking   prevents dynamical breaking
of chiral symmetry.  Consequently, one would obtain a
confining but chirally symmetric phase within the quarkyonic matter
 \cite{GW}.
 To this end one needs a quark that is confined but at the same time
 its Green function is chirally symmetric. Is it possible?
 
There is no way to answer today this question within QCD itself. What
can be done is to clarify the issue whether it is possible or not
in principle. If possible,  a key question
is about the physical mechanism that could be behind
such a phase. Then  to address this question one needs a model
that is manifestly confining, chirally symmetric and provides dynamical
breaking of chiral symmetry. Such a model does exist. The answer
to the question above is "yes", at least within the model.

\section{Confining and chirally symmetric model}

We will use the simplest possible model that is manifestly
confining, chirally symmetric and guarantees dynamical breaking
of chiral symmetry in a vacuum \cite{Y,ADLER}. It is assumed
within the model that the only gluonic interaction between quarks
is a linear instantaneous potential of the Coulomb type.
(This model can
be considered as a 3+1 dim generalization of the 't Hooft model \cite{HOOFT}.
In the 't Hooft model, that is the large $N_c$ QCD in 1+1 dimensions, the
only interaction between quarks is a linear confining potential
of the Coulomb type.)
Such  potential in 3+1 dimensions
is a main ingredient of the Gribov-Zwanziger scenario in Coulomb
gauge \cite{GZ} and is indeed observed in Coulomb gauge variational 
calculations \cite{SW} as well as in Coulomb gauge lattice simulations
\cite{V}. 

A key point is that the quark Green function (that is  a solution of the gap
equation in a vacuum) contains not only the chiral symmetry breaking part $A_p$,
but also the manifestly chirally symmetric part $B_p$: 

\begin{equation}
\Sigma(\vec p) =A_p +(\vec{\gamma}\cdot\hat{\vec{p}})[B_p-p].
\label{SE} 
\end{equation}

The linear potential requires the infrared regularization. Otherwise all loop
integrals are infrared divergent. All observable color-singlet quantities
are finite and well defined in the infrared limit (i.e., when the infrared cutoff approaches
zero). These are hadron masses, the quark condensate, etc. In contrast,  all color-nonsinglet quantities 
are divergent. E.g., single quarks have infinite energy and consequently are removed from the spectrum.  
This is a manifestation of confinement within this simple model.

Given a quark Green function obtained from the gap equation,
one is able to solve the Bethe-Salpeter equation for mesons.
A very important aspect of this model is that it exhibits the
effective chiral restoration in hadrons with large $J$ \cite{WG,G3,NB}.
This is because  chiral symmetry breaking is important only
at small momenta of quarks. But at large $J$ the centrifugal
repulsion cuts off the low-momenta components in  hadrons  
 and consequently the hadron wave function and its mass are
insensitive to the chiral symmetry breaking in the vacuum. The
chiral symmetry breaking in the vacuum represents only a tiny
perturbation effect: Practically the whole hadron mass comes
from the chiral invariant dynamics. This  explicitly
demonstrates that it is possible to construct hadrons in such a way
that their mass origin is not the quark condensate. 
If so, it is clear apriori that there are good chances to obtain a confining
but chirally symmetric matter  within this model.

Now we want to see what will happen with confinement and chiral
symmetry at zero temperature and large density. There are no
quark loops within this model and consequently Debye screening
of the confining potential is absent. Confinement persists up
to arbitrary large density. Will it be possible to restore chiral
symmetry at some density?

This 3+1 dim model is complicated enough and it is not
possible to solve it exactly for a dense baryonic matter. What
can be done is a kind of a mean-field solution. To obtain
such a solution we need an additional assumption. Namely, we
assume that both rotational and translational invariances are
not broken in a medium. This implies that we assume a liquid
phase. 

This assumption is rather important and it is worth to discuss
its relevance. We know that  within the $N_c=3$
QCD both translational and rotational invariances are not broken
in the one nucleon system.
We also know that the nuclear matter is a liquid, i.e., the 
translational and
rotational invariances are not broken. 

This real life situation
is drastically different from exactly solvable 1+1 dimensional
theories (the 't Hooft model
and the Gross-Neveu model) \cite{TH}. In the latter cases the
translational invariance is broken both for a one-nucleon and
baryonic matter solutions and one obtains a chiral spiral with
inhomogeneous chiral condensate. The baryonic matter is  in
a crystal phase. In contrast, we  assume that in the real life a dense 
baryonic matter is a liquid, like standard nuclear matter. Such
 assumption is also supported by the fact that at $N_c=3$ and very
high densities the relevant phase is a color-superconducting matter,
where the system is also a (ideal) liquid. It is difficult
to imagine a phase diagram where at $T=0$ a liquid - crystal - liquid sequence
of phases with increasing density would exist. Hence we  assume
that at zero temperature we always have a liquid, all the
way up to very high density. Obviously, properties of
this liquid should be quite different at different densities.
Such assumption implies that both the rotational and
translational invariances are not broken in a dense matter.

Given  unbroken translational and rotational invariances
 and assuming that confinement survives
up to rather large density we can now try to answer the question
about possible existence of confining but chirally symmetric
phase \cite{GW}.

Imagine  that we have a dense baryon (quarkyonic) matter consisting of 
overlapping baryons
with a well defined quark Fermi sea and the quark 
Fermi momentum is $P_f$. At the same time the interquark linear potential is not
yet screened (in the  large $N_c$ limit it is not screened at any density). 
In order to 
understand what happens with  chiral symmetry
we have to solve a gap equation for a probe quark with momentum
above $P_f$.
  All intermediate 
quark levels below $P_f$ are Pauli blocked and do not contribute to the gap 
equation. Consequently, at sufficiently
large $P_f$ a chiral restoration phase transition happens, see Fig. 1.
Chiral symmetry is restored like in the NJL model, because there is not 
available phase space
in the gap equation to create a nontrivial solution with broken 
chiral symmetry. This required
phase space is removed by the Pauli blocking of  levels with positive energy. 
The standard 
quark-antiquark condensate of the vacuum vanishes.

\begin{figure}[h]
    \centering
        \includegraphics[width=0.45\textwidth]{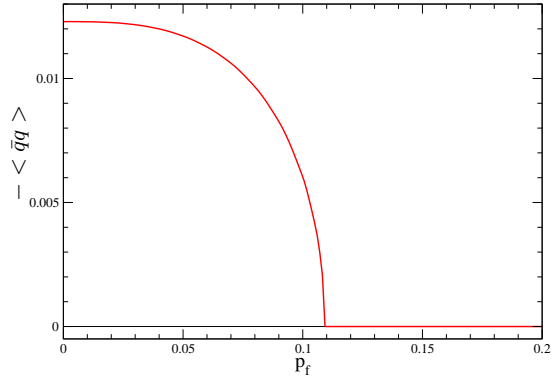}
\caption{Chiral restoration phase transition in a dense quarkyonic
matter with unbroken translational and rotational symmetries.}
\label{Fig.1}
\end{figure}

Above the chiral restoration phase transition the 
chiral symmetry breaking
part of the quark self-energy identically vanishes, $A_p=0$. 
What crucially distinguishes this confining model from the
nonconfining models like NJL, is that the quark Green function
contains also a chirally symmetric part, $B_p$. This
chirally symmetric  quark self-energy, $B_p$,  
does not vanish both below and above the chiral restoration phase
transition and is  infrared-divergent,
like in vacuum. This means that even in the chirally restored regime the 
single quark energy is infinite
and a single quark is removed from the spectrum. 
This infrared divergence is exactly canceled, however,
in any color-singlet excitation of baryonic or mesonic type. 
Consequently, a spectrum of excitations
consists of a complete set of all possible chiral multiplets. 
Energies of these excitations  are finite
and well defined quantities. A mass of this confining matter 
is chirally symmetric
and comes  from the chiral invariant dynamics.
 
Such  confined phase with restored chiral symmetry can be viewed as a system of chirally
symmetric baryons that are in a strong overlap. Confining gluonic
fields are not screened, but quarks can move not only within each individual baryon,
but also within the matter by hoping from one baryon to another. 
So one cannot say to which specific baryon a given quark belongs.  
However, it would be a mistake to consider these quarks as free
particles. They are colored and are subject to strong gluonic confining fields.
 Their dispersion
law is a complicated one (it contains the infrared divergent term), 
by far not as for free particles. This picture is different from the
naive perlocation picture, where one thinks that due to perlocation 
of baryons the quarks are free.

An interesting question is what happens near the Fermi surface of such  dense
confining matter with vanishing quark-antiquark condensate. Could be there 
 some
surface phenomena like chiral density waves  \cite{Ru}? These chiral density
waves have been derived so far as an instability of the quark Fermi sea
(with free unconfined quarks) due to a gluon exchange force between
quarks and quark holes with large momenta near the Fermi surface. However,
in the confining baryonic mode the relevant degrees of freedom near the Fermi
surface are color singlet baryons. In such case all lowest
excitations are of the
baryon - baryon hole type (i.e., of three quarks - three quark holes type),
because it costs no energy to excite a baryon from the Fermi surface
to the next not occupied level.
Here the gluon-exchange force in the quark - quark hole pairs is simply absent
and the chiral density waves cannot be formed. The quark - quark hole excitations
with the color - exchange force between the quark and the quark hole
would correspond to intrinsic excitations of baryons near the Fermi
surface which costs a lot of energy. 

\section{What order is the chiral restoration phase transition?}

A very interesting question is about the order of the phase
transition from the confining matter with broken chiral symmetry
to the confining matter with restored chiral symmetry at low temperatures. 
Within the above mentioned model 
this question can be answered by microscopical calculations.
However, such  model answer to this question is 
of little relevance to real QCD with $N_c=3$.

The point is that in vacuum the hadron spectrum within the model above can be
considered as  a  set of linear chiral multiplets and the chiral
partners within each multiplet are split by the chiral symmetry
breaking effects \cite{GW}. Hence, there is a one - to - one correspondence
between the chiral partners. Consequently, across the phase transition
there is a continuous transformation of hadrons  from the
broken chiral symmetry phase to the phase with restored chiral symmetry
and the phase transition is  of the second order.

However, in real QCD it is by far not so. Chiral
symmetry is  strongly broken in the lowest lying hadrons,
so there are not chiral partners to hadrons like nucleon or $\rho$-meson.
(It would be possible to assign approximate chiral partners
only if the chiral symmetry breaking effect were weak). This
can be directly seen from the lowest lying spectroscopy in vacuum:
there is not a one - to - one correspondence of positive and
negative parity hadrons in the low  lying part of the spectrum.
Hence, it is not possible to arrange all low lying
hadrons into approximate linear chiral multiplets. Only for
the high lying hadrons such a correspondence can be seen.

This question can be also investigated on the lattice. The chiral
content of the $\rho$-meson has been studied in dynamical
simulations \cite{GLL}. It turned out that the physical $\rho$-meson is
a strong mixture of two chiral representations and it is
not possible to assign  the lowest $a_1$  and
the $h_1$ mesons as rho's chiral partners. For the nucleon
the chiral symmetry breaking effect is probably even stronger
and the nucleon is a strong mixture of several different chiral
representations. 

In the chirally symmetric phase with
confinement {\it all} hadrons must be arranged into chiral
multiplets. Then it is not possible to connect continuously
the lowest lying hadrons in the Nambu-Goldstone mode, that
have no chiral partners, with  hadrons in the Wigner-Weyl
mode, which are all arranged into exact chiral multiplets. The
phase transition is then necessarily discontinuous, i.e.,
of the first order.

A possible schematic phase diagram is shown on Fig. 2.

\begin{figure}[h]
    \centering
        \includegraphics[width=0.45\textwidth]{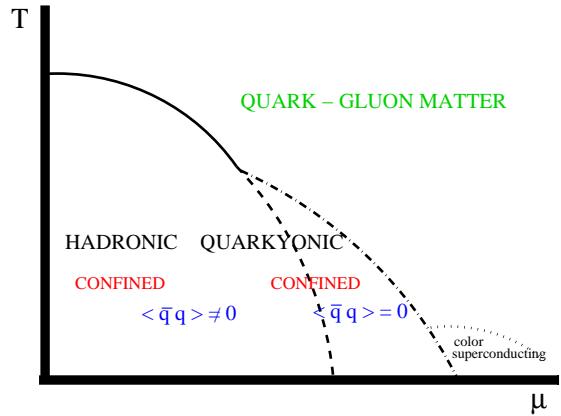}
\caption{ A schematic phase diagram.}
  \label{Fig.2}
\end{figure}

\medskip

{\bf Acknowledgements}

Support of the Austrian Science
Fund through the grant P21970-N16 is acknowledged.

\end{document}